\documentclass[prl,aps,epsf,showpacs,twocolumn]{revtex4-1}
\usepackage{times}
\usepackage{graphicx}
\usepackage{float}
\usepackage{latexsym,amsmath,amssymb,bm,euscript}
\usepackage{color}
\usepackage{subfigure}
\usepackage{epstopdf}
\usepackage[colorlinks=true,linkcolor=blue,citecolor=blue,urlcolor=blue]{hyperref}
\usepackage{hyperref}
\usepackage{ulem}
\usepackage[dvipsnames]{xcolor}
\usepackage{amsmath}
\def\ET{{$\kappa$-(ET)$_2$Cu$_2$(CN)$_3$}}
\def\dmit{{EtMe$_3$Sb[Pd(dmit)$_2$]$_2$}}

\begin{document}

\title{Fractionalized Excitations Revealed by Entanglement Entropy}
\author{Wen-Jun Hu$^{1,2}$} 
\author{Yi Zhang$^{3}$} 
\author{Andriy H. Nevidomskyy$^{2}$}
\author{Elbio Dagotto$^{1,4}$}
\author{Qimiao Si$^{2}$}
\author{Hsin-Hua Lai$^{2}$}
\affiliation{
$^1$ Department of Physics and Astronomy, University of Tennessee, Knoxville, Tennessee 37996, USA\\
$^2$ Department of Physics and Astronomy \& Rice Center for Quantum Materials, Rice University, Houston, Texas 77005, USA\\
$^3$International Center for Quantum Materials, Peking University, Beijing, 100871, China\\
$^4$ Materials Science and Technology Division, Oak Ridge National Laboratory, Oak Ridge, Tennessee 37831, USA
}

\begin{abstract}
Fractionalized excitations develop in many unusual many-body states such as quantum spin liquids, disordered phases that cannot be described using any local order parameter. Because these exotic excitations correspond to emergent degrees of freedom, how to probe them and establish their existence is a long-standing challenge. We present a general procedure to reveal the fractionalized excitations using real-space entanglement entropy in critical spin liquids that are particularly relevant to experiments. Moreover, we show how to use the entanglement entropy to construct the corresponding spinon Fermi surface. Our work defines a new pathway to establish and characterize exotic excitations in novel quantum phases of matter.
\end{abstract}

\maketitle

{\it Introduction ---}
A hallmark of strongly correlated systems is the emergence of novel degrees of freedom at low energies from strong correlations. A prototype case is fractionalized excitations -- fundamentally different from excitations in weakly interacting limit -- such as spinons in Herbertsmithite ZnCu$_3$(OH)$_6$Cl$_2$~\cite{kagome} and YbMgGaO$_4$~\cite{li2015gapless,paddison,shen2018}. A particularly intriguing possibility arises in quantum spin liquids because their emergent fermionic excitations can form a Fermi surface in momentum space, rendering the properties of these insulators akin to those of conventional metals. The two-dimensional (2D) triangular lattice-based organic compounds \dmit~and~\ET~\cite{Shimizu03,SYamashita08,MYamashita09,Itou10} are among the most famous candidate materials believed to host such a critical spin liquid (CSL) with an emergent spinon Fermi surface (SFS)~\cite{LeeNagaosaWen, Savary_QSLreview}. A four-spin ring exchange is needed to describe these materials~\cite{ringxch,Fisher2011,Patrick2018}. An outstanding challenge is how to demonstrate and reveal the presence of fractionalized fermionic excitations, particularly with regards to the SFS.

On the theory side, it has been proposed to study the emergent SFSs in CSLs through the singular peaks in the spin structure factor (SSF) -- those that arise from real-space power-law decaying spin correlations -- which can be related to the locations of the SFS~\cite{LeeNagaosaWen}. Using this procedure, recent density matrix renormalization group (DMRG) results reported the possible SFS of the spin-$1/2$ model on a triangular lattice with a four-spin ring exchange~\cite{Fisher2011,Patrick2018} and in the Kitaev model on a honeycomb lattice~\cite{Patel12199}. However, it is still difficult to reconstruct the actual shape of the SFS through the DMRG results of the SSFs based on small system sizes.

An alternative quantity to describe long-range entangled states is the entanglement entropy (EE)~\cite{QE_RMP}, such as the von Neumann EE and the Renyi EE (REE), which are obtained from reduced density matrix of a subsystem by tracing out the degrees of freedom outside this subsystem. The EE plays an important role in several fields, ranging from quantum information to condensed matter physics~\cite{Amico_RMP}, and has been measured experimentally~\cite{Greiner_EE_nature}. It is believed that the EE of the ground states in most local Hamiltonians satisfies the ``EE area law''~\cite{Eisert_RMP}: when a system is divided into subsystems, the EE is proportional to the area of the boundary between the two subsystems at the leading order.

Violations of the EE area law do exist in various cases. In one dimension, they are found in several quantum critical systems~\cite{Calabrese2009}. In higher dimensions, these violations are associated with the presence of a SFS in momentum space. The most well-known examples are the ground states of free fermions with Fermi surfaces~\cite{Wolf, GioevKlich}, where the violation is logarithmic, i.e. the EE is proportional to the surface area multiplied by a factor that grows logarithmically with the subsystem size. Intriguingly, the EE in these noninteracting systems takes the Widom formula~\cite{Wolf, GioevKlich, Swingle2010}, where the coefficient of the leading term in the dependence of EE on the subsystem size captures the geometric information of the Fermi surface and that of the subsystem. For gapless electronic systems, calculations perturbative in the interactions~\cite{dingprx12} show that such a violation retains the same form as that of a free Fermi gas. Recently, it has been suggested that the EE associated with the composite Fermi liquid phase of the half-filled Landau level ($\nu = 1/2$) is also described by the Widom formula~\cite{Mishmash2016}. By contrast, for frustrated strongly correlated electrons, as in Hubbard models, or spin systems, as in Heisenberg models, all with a possible emergent SFS, the EE has not been much explored~\cite{Lai_EE4EBL,Zhang11}.

Our present work goes beyond previous efforts and provides a generic procedure to reconstruct the geometry of the emergent SFS. We present the first variational Monte Carlo (VMC) study of the EE to test the conjectured Widom formula for strongly correlated systems. Employing a widely discussed example of a CSL with an emergent SFS, we introduce a direct probe of emergent fractionalized excitations using the real-space EE together with examining the singularity of the SSF. Remarkably, we show that the leading order of the EE has the form of the Widom formula multiplied by a previously unknown factor of $2$. This numerical factor captures the presence of two free gapless modes associated with two flavors of spinons. From this formula, we provide the basis~\cite{Lai_EEreconstruct} for a systematic methodology to explicitly reconstruct the emergent SFS geometry. We remark that using the SSF or EE individually only allows you to test the existence or not of fractionalized excitations (i.e. a "yes" or "no" answer), but a combined methodology is necessary to recover the full shape of the SFS. We also remark that we employ VMC only for simplicity: our methodology can be used if other techniques are employed, such as the quantum Monte Carlo technique or DMRG. With the only caveat that it is advisable to employ several trial states to search for self-consistency to remove the bias uncertainty intrinsic of variational procedures, our procedure is quite generic.

\begin{figure*}
   \centering
   \includegraphics[width=0.8\linewidth]{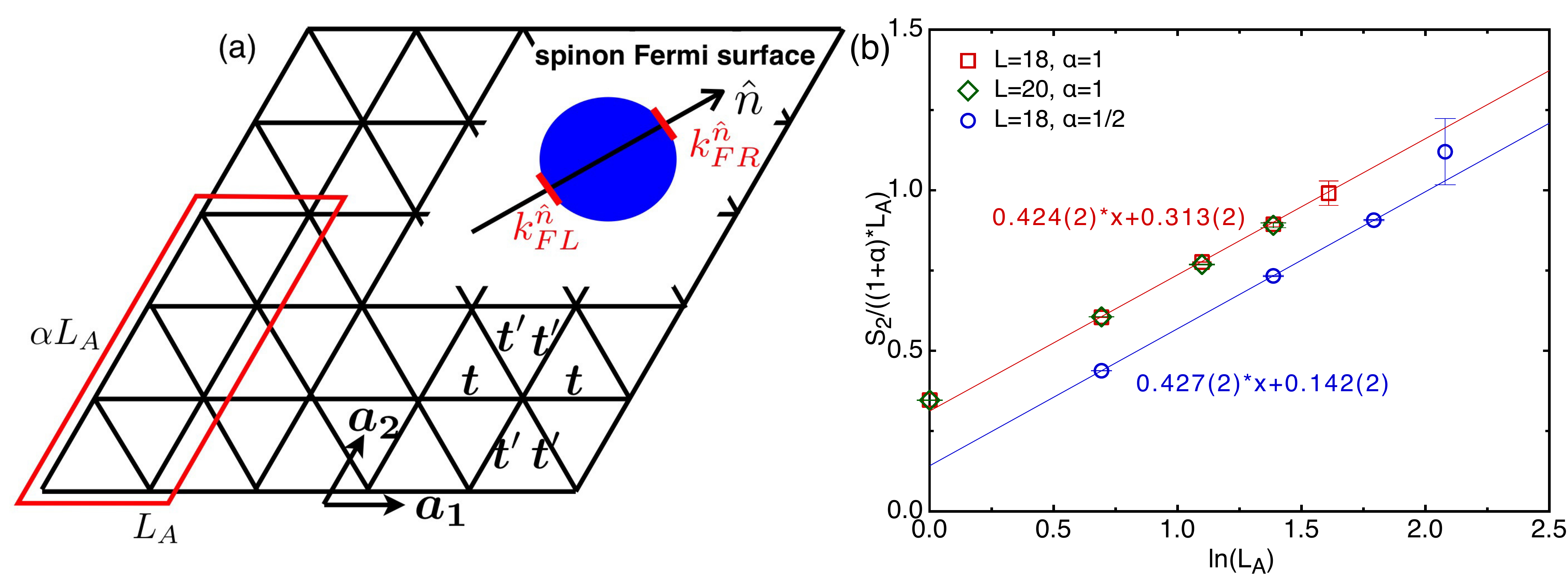}
   \caption{(a) Illustration of the subsystem geometries used to obtain the REE on the triangular lattice. In the simplest case, we consider a subsystem (illustrated in the left bottom part) consisting of $L_A \times \alpha L_A$ sites along $\bm{a}_1 = (1, 0) $ and $\bm{a}_2 =(1/2, \sqrt{3}/2)$ directions. The top right portion illustrates the emergent SFS. For a direction $\hat{n}$, there are two Fermi patches perpendicular to $\hat{n}$ that define momenta $k^{\hat{n}}_{FR/L}$ for the right or left Fermi patches. For an emergent surface with an inversion center -- natural for a system with time-reversal symmetry or inversion symmetry -- the length of the difference between $\bm{k}^{\hat{n}}_{FR}$ and $\bm{k}^{\hat{n}}_{FL}$ gives its cross section along $\hat{n}$. (b) REE, $S_2$, for different subsystem geometries using the isotropic Gutzwiller-projected WF. Based on Eq.~\eqref{Eq:Widom_simplified}, we plot $ S_2/( (1+\alpha)L_A)$ vs $\ln (L_A)$. The slopes of the lines give the prefactor of the leading term of the REE. We fix the whole system size and choose the subsystem size to be $L_A \times \alpha L_A$ with $\alpha=1$ and $1/2$. The data on $L=18$ with $\alpha=1$ (red squares) are consistent with Ref.~\cite{Zhang11}. We can clearly observe the proportionality $S_2 \propto (1+\alpha)L_A \ln L_A$.}\label{Fig:tri_iso_sim}
\end{figure*}

{\it Entanglement entropy and Widom formula ---}
We will first provide robust numerical evidence for the validity of the Widom formula in a CSL with emergent SFS. A typical ground-state wave function (WF) to represent the possible CSL on a triangular lattice~\cite{Shimizu03,SYamashita08,MYamashita09,Itou10} is the Gutzwiller projected Slater determinant: $|\psi\rangle=\mathcal{P}_{G}|\psi_{0}\rangle$, where the Gutzwiller projector $\mathcal{P}_{G}=\prod_{i}(1-n_{i\uparrow}n_{i\downarrow})$ forbids double occupation on each site, and $|\psi_{0}\rangle$ is the ground state of the mean-field Hamiltonian on the triangular lattice ${\cal H}_{\rm MF} = \sum_{\langle i,j\rangle,\sigma} t_{ij}c^{\dag}_{i,\sigma}c_{j,\sigma} + h.c.$. The Gutzwiller projector $\mathcal{P}_{G}$ is crucial to avoid a trivial Fermi surface of real electrons, while still allowing a possible SFS. This variational WF is known to be accurate for the quasi-1D $J_1-J_2$ spin-$1/2$ chain with four-spin exchanges~\cite{Sheng09}, providing a reasonable starting point for our effort. We begin by considering an isotropic system with a total number of sites $N_s=L\times L$ [$t'=t$ in Fig.~\ref{Fig:tri_iso_sim}(a)].

Based on the Widom formula, the REE associated with a subsystem consisting of $L_A \times \alpha L_A$ sites along the $\bm{a}_1\equiv (1,0)$ and $\bm{a}_2\equiv (1/2, \sqrt{3}/2)$ directions (lattice constant $a\equiv 1$), as illustrated in the bottom left portion of Fig.~\ref{Fig:tri_iso_sim}(a), can be concisely expressed as follows (derivations in Supplemental Material~\cite{suppl})
\begin{eqnarray} \label{Eq:Widom_simplified}
S_{2} &\dot{=}& \frac{c_{eff}}{8\pi} (1+\alpha)A_{sf} L_A \ln L_A \nonumber \\
&=& \frac{c_{eff}}{8\pi} (1+\alpha) \left| \bm{k}^{\hat{n}}_{FR} - \bm{k}^{\hat{n}}_{FL} \right| L_A \ln L_A,
\end{eqnarray}
where $\dot{=}$ means the leading logarithmic contribution in REE, $\alpha$ represents the ratio between the linear length of the subsystem ($L_A$) and that of the whole system ($L$), \textit{i.e.} $\alpha = L_A / L$, and $c_{eff}$ is effectively the number of free gapless modes in the low-energy limit. Additionally, $A_{sf}$ refers to the cross section of the SFS, which is determined by the span in the momenta between right or left moving patches ($\bm{k}^{\hat{n}}_{FR/L}$) of the SFS along any particular observation direction $\hat{n}$. This is illustrated in the top right portion of Fig.~\ref{Fig:tri_iso_sim}(a), where the emergent SFS is expected to be circular.

We have carried out the VMC simulations on the triangular lattice with the whole system size fixed to be $L \times L$, with $L$ up to $20$. We calculated the REE associated with a subsystem of $L_A \times \alpha L_A$ sites, where both $L_A$ and $\alpha L_A$ are less than or equal to $L/2$. The resulting REE vs $L_A$ is plotted in Fig.~\ref{Fig:tri_iso_sim}(b), which shows that $S_2/ ((1+\alpha)L_A)$ \textit{vs.} $\ln L_A$ has the same slope for different choices of $\alpha$ within error bars. The proportionality $S_2 \sim  (1+\alpha)L_A\ln L_A$ provides direct evidence that the REE of the CSL studied here satisfies the Widom formula Eq.~\eqref{Eq:Widom_simplified}. The slope in Fig.~\ref{Fig:tri_iso_sim}(b) gives the value of the combined variable $c_{eff} A_{sf} = c_{eff} \left| \bm{k}^{\hat{n}}_{FR} - \bm{k}^{\hat{n}}_{FL}\right|$. In order to pin down the explicit formula for the REE of a CSL, additional information is needed to determine the values of $c_{eff}$ and $A_{sf}^{\hat{n}}$ separately, as addressed next.

\begin{figure*}
   \centering
   \includegraphics[width=0.8\linewidth]{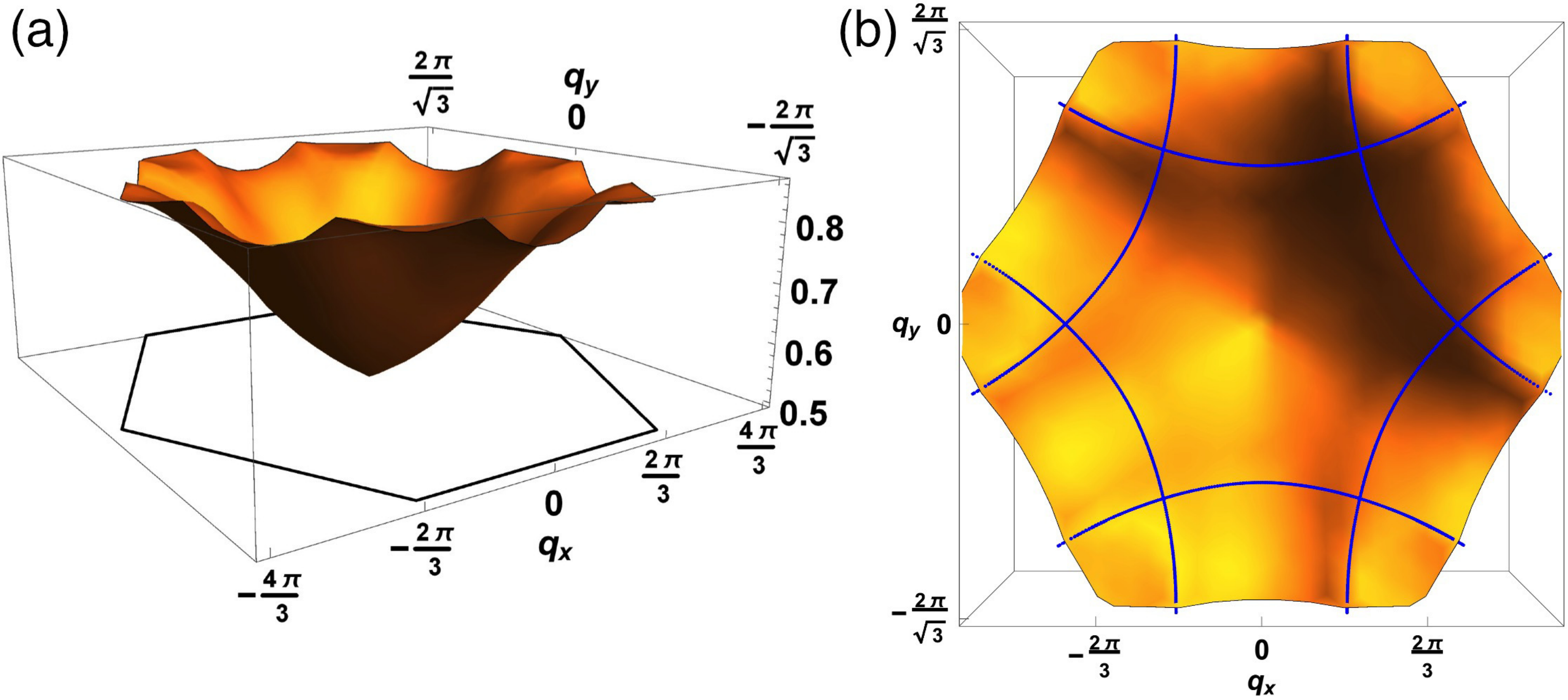}
   \caption{The 3D plot of SSF for the isotropic case. (a) Side view of the SSF within the BZ for the isotropic case (black hexagon represents the BZ), employing a triangular 30$\times$30 cluster. There is a sharp and singular peak at ${\bf q}=0$, which corresponds to the uniform real-space power-law decaying behavior. The much weaker singular lines near the boundary of the BZ correspond to the oscillating real-space behavior caused by the presence of the SFS. (b) Top view of the SSF. The blue lines are fitting results that are seen to match well with the singular lines for the SFS in the SSF on a triangular lattice consisting of $30\times 30$ sites. The details of the fitting method are in Supplemental Material~\cite{suppl}.}
   \label{Fig:sq_uni}
   \end{figure*}

{\it Spin structure factor ---}
Using the VMC described earlier, we calculated the SSF $D_{\bm{q}} \equiv \frac{1}{N_s} \sum_{i,j} \langle {\bf S}_{i}\cdot {\bf S}_{j} \rangle e^{i{\bf q}\cdot({\bf r}_i-{\bf r}_j)}$ with the spin operator ${\bf S}_{i}=\sum_{\sigma,\sigma'} {{1}\over{2}}c^\dagger_{i \sigma} \boldsymbol{\sigma}_{\sigma,\sigma'} c^{\phantom\dagger}_{i \sigma'}$. It is known that for an arbitrary observation direction $\hat{n}$, $D_{\bm{q}}$ should show singular peaks at $\bm{q} = \bm{0}$ and~$\bm{k}^{\hat{n}}_{FR} - \bm{k}^{\hat{n}}_{FL}$, which are associated with forward and backward scattering processes. The information of $D_{\bm{q}}$ can be used to determine the cross section of the emergent SFS whose surface unit vector is perpendicular to $\hat{n}$, \textit{i.e.}, $A_{sf}^{\hat{n}} = \left| \bm{k}^{\hat{n}}_{FR} - \bm{k}^{\hat{n}}_{FL}\right|$~\cite{Sheng09}. In the isotropic case, $A_{sf}^{\hat{n}} = A_{sf}$ is independent of the direction.

In Fig.~\ref{Fig:sq_uni} we show the numerical data for the SSF on a triangular lattice with $30 \times 30$ sites. Figure~\ref{Fig:sq_uni}(a) gives a 3D side view of the SSF in the Brillouin zone (BZ), denoted by the black hexagon, where we can see a sharp singular point at $\bm{q} = 0$ and weaker singular lines on the surface whose locations are theoretically suggested to be $\bm{q}=\bm{k}^{\hat{n}}_{FR} - \bm{k}^{\hat{n}}_{FL}$. Figure~\ref{Fig:sq_uni}(b) shows the 3D top view of $D_{\bm{q}}$. In the present finite-size calculations, the singular lines on the 3D $D_{\bm{q}}$ surface are more clearly revealed near the BZ boundary, while the weaker singular lines inside the BZ are masked by the sharper singular point at $\bm{q}=0$. From Fig.~\ref{Fig:sq_uni}(b), we can determine the location of the full singular lines by fitting $\bm{k}^{\hat{n}}_{FR} - \bm{k}^{\hat{n}}_{FL}$~\cite{Lai_EEreconstruct}, which allows us to extract the (average) cross sections of the emergent SFS to be $5.24\pm 0.05$. When this value for the cross section is combined with the slopes of the normalized REE {\it vs.} $\ln (L_A)$ shown in Fig.~\ref{Fig:tri_iso_sim}(b), we obtain $c_{eff} \simeq 2.01\pm 0.02$. This value indicates the presence of two free gapless modes for each ``independent" 1D patch in the low-energy limit \cite{SSLee2009}, so it should be universal for all shapes of convex critical Fermi surfaces. If we introduce anisotropy into the system, $c_{eff}$ should remain the same.

\begin{figure*}[t]
   \centering
   \includegraphics[width=0.8\linewidth]{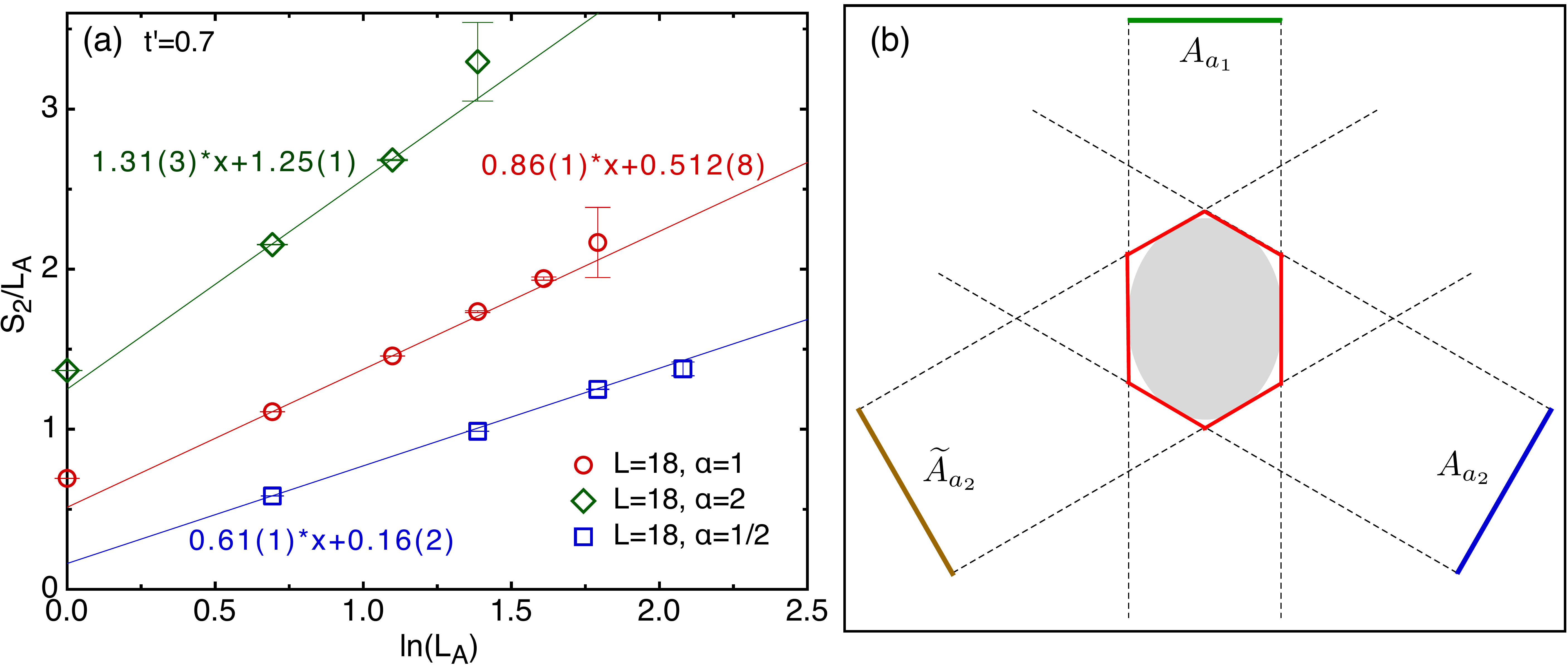}
   \caption{(a) EE in the presence of anisotropy along the zigzag bonds. We fix the whole system size to be $18\times 18$ and choose the subsystem size to be $L_A \times \alpha L_A$ with $\alpha = 1/2$ (blue squares), $1$ (red circles), and $2$ (green diamonds) to extract the REE. (b) Reconstruction of the SFS with an inversion center. Based on the REE results in (a), we can obtain the cross sections of the SFS projected onto $\bm{a}_{1/2}$ axis, $A_{\bm{a}_{1/2}}$ (green and blue lines). Due to the presence of an inversion center, we can draw an inverted partner of $A_{\bm{a}_2}$ denoted as $\tilde{A}_{\bm{a}_2}$ (brown line). The dashed lines are perpendicular to $A_{\bm{a}_{1/2}}$ and $\tilde{A}_{\bm{a}_2}$, and connecting all the intersections of the dashed lines gives the shape of the SFS in the lowest order. The light-gray ellipse is the SFS obtained by extracting $\bm{k}^{\hat{n}}_{FR} - \bm{k}^{\hat{n}}_{FL}$ from the SSF.}\label{Fig:ee_ani07}
\end{figure*}

{\it Visualizing emergent spinon Fermi surface ---}
The explicit formula for the EE obtained above can be used to reveal the emergent SFS directly. For an isotropic system, since the shape of an emergent SFS is circular, and its diameter can be extracted once the REE is calculated. To address a more general case, we focus on a triangular lattice system with anisotropy. Specifically, we consider a Gutzwiller-projected WF with hopping amplitudes $t$ along each ladder ($\pm \bm{a}_1$ directions) and $t'$ along the zigzag directions ($\pm \bm{a}_2$ and $\pm (\bm{a}_1 - \bm{a}_2)$) that couple different ladders as shown at the bottom right of Fig.~\ref{Fig:tri_iso_sim}(a). For an illustration, we use $t'/t =0.7$ to obtain the REE associated with the subsystems. 

Because of numerical and computational time limitations, below we choose three subsystem geometries to obtain the REE and thereby construct the anisotropic SFS. Specifically, we calculate the REE for a subsystem with $L_A \times \alpha L_A$ sites, where we consider ratios $\alpha = 1/2,~1,~2$. The REE results in these systems are shown in Fig.~\ref{Fig:ee_ani07}(a). Setting $c_{eff} = 2$ for the present anisotropic system the formula for REE becomes:
\begin{eqnarray}
S_2 \dot{=} \frac{1}{4\pi} \left( \alpha A_{\bm{a}_2}+ A_{\bm{a}_1}\right) L_A \ln (L_A),
\end{eqnarray}
where $A_{\bm{a}_{1 or 2}}$ represents the cross sections of the SFS projected onto the $\bm{a}_{1 or 2}$ axis. We can write down three equations, corresponding to $\alpha=1/2$, $1$, and $2$, respectively: ({\it i}) $A_{\bm{a}_1} + A_{\bm{a}_2} = 4\pi *0.86 \pm 0.01$, ({\it ii}) $A_{\bm{a}_1} +  A_{\bm{a}_2}/2 = 4\pi * 0.61 \pm 0.02$, and ({\it iii}) $A_{\bm{a}_1} + 2 A_{\bm{a}_2} = 4\pi * 1.31 \pm 0.03$. We can choose any two out of the three equations to obtain the values of $A_{\bm{a}_{1 or 2}}$. Since there are three choices, we can obtain three numerical approximations for $A_{\bm{a}_{1 or 2}}$, which we average over to reduce the statistical error. We find that $A_{\bm{a}_1} \simeq 1.53 \pi$ and $A_{\bm{a}_2} \simeq 1.89 \pi$, based on which the shape of the emergent SFS can be constructed as illustrated in Fig.~\ref{Fig:ee_ani07}(b). The green and the blue lines represent the cross sections of $A_{\bm{a}_1}$ and $A_{\bm{a}_2}$. Since there is an inversion center for the emergent surface in momentum space~\cite{Lai_EEreconstruct}, once $A_{\bm{a}_2}$ is known, we can draw its inverted partner, denoted as $\tilde{A}_{\bm{a}_2}$ (brown line) in Fig.~\ref{Fig:ee_ani07}(b). The dashed lines are perpendicular to $A_{\bm{a}_{1/2}}$ and $\tilde{A}_{\bm{a}_2}$, respectively. Connecting all the intersections of the dashed lines results in the red hexagonal shape, which provides the leading-order approximation to the shape of the emergent SFS. In principle, we can improve the accuracy of the shape if we perform more (time consuming) REE calculations using different subsystem geometries~\cite{Lai_EEreconstruct}.

For comparison, we also show the shape of the SFS in Fig.~\ref{Fig:ee_ani07}(b) (light-gray ellipse) obtained by extracting $\bm{k}^{\hat{n}}_{FR} - \bm{k}^{\hat{n}}_{FL}$ from the SSF. The exact numerical results for the SSF are shown in Supplemental Material.~\cite{suppl} The emergent SFS reconstructed from the REE results is quite consistent with the light-gray ellipse in Fig.~\ref{Fig:ee_ani07}(b), which provides additional support for our procedure. With (costly) additional values of $\alpha$ our results will be even closer to the ellipse. We remark that in strongly correlated systems, where analytical methods are difficult to use and numerical simulations only can be performed on small clusters, it may be difficult (or sometimes impossible) to determine the locations of $\bm{k}^{\hat{n}}_{FR} - \bm{k}^{\hat{n}}_{FL}$ and thus the here proposed EE probe becomes the only practical procedure, exhibiting its unique value. From this overarching perspective, the present work builds up a foundation for using the EE to probe emergent SFSs in general cases.

{\it Conclusion and outlook ---}
In this work, we examined the entanglement properties of a CSL with an emergent SFS. Numerically, we have proved the validity of a generalized Widom formula Eq.~\eqref{Eq:Widom_simplified}~\cite{suppl} for this type of strongly correlated systems. Based on this formula, we provide a general procedure to reveal and construct the shape/size of emergent SFSs, by examining the singularity of the SSF and the real-space EE. This is an advance over previous efforts that relied on the singular peaks in the SSFs to locate the SFS by DMRG, because using only the latter the whole shape of the SFS cannot be obtained. In addition, we have obtained the universal factor $c_{eff}=2$ that describes two free gapless modes in a CSL employing robust numeral calculations, without ``guessing'' this value in advance.

The current work can be straightforwardly generalized to CSLs of higher-spin ($S\geq1$) systems. Of particular interest is the 6H-B phase of $S=1$ Ba$_3$NiSb$_2$O$_9$~\cite{Zhou2011, Mendels2016} that was recently suggested to realize a CSL with three flavors of fermionic spinons, forming a large SFS~\cite{Bieri2017}. From our perspective, it is always possible to write down a Gutzwiller-projected WF of three flavors of fermions to represent the $S=1$ CSL. Based on the results presented here, we conjecture that the leading EE in this case also satisfies the Widom formula, but with $c_{eff} = 3$. Finally, our work points to new prospects for deepening the understanding of correlated systems such as heavy-fermion materials, in which the nature of quantum spins and Fermi surface plays a crucial role~\cite{siheavy}. Examining the quantum entanglement properties promises a conceptually new way of elucidating their quantum phases and criticality.

{\it Acknowledgments. ---} We thank Federico Becca, Kun Yang, and Lesik Motrunich for helpful discussions. E.D. and W.-J.H. were supported by the U.S. Department of Energy (DOE), Office of Science, Basic Energy Sciences (BES),  Materials  Science  and  Engineering  Division. The work was supported in part by the NSF Grant No.\ DMR-1920740 and the Robert A.\ Welch Foundation Grant No.\ C-1411 (W.-J.H., H.-H.L. and Q.S.), a Bethe Fellowship at Cornell University (Y.Z.), the NSF Grant No.\ DMR-1350237 (W.-J.H, H.-H.L. and A.H.N.), a Cottrell Scholar Award from the Research Corporation for Science Advancement, and the Robert A.\ Welch Foundation Grant No.\ C-1818 (A.H.N.), and a Smalley Postdoctoral Fellowship of the Rice Center for Quantum Materials (H.-H. L.). A.H.N. acknowledges the hospitality of the Aspen Center for Physics, which is supported by National Science Foundation Grant No. PHY-1607611. 
The majority of the computational calculations have been performed on the Extreme Science and Engineering Discovery Environment (XSEDE) supported by NSF under Grant No.\ DMR160057. Most of the numerical calculations have been done by W.-J.H. and H.-H.L. while at Rice University. 

\bibliography{GW_EE}
\end{document}


\title{Supplemental Material for Fractionalized Excitations Revealed by Entanglement Entropy}
\author{Wen-Jun Hu$^{1,2}$} 
\author{Yi Zhang$^{3}$} 
\author{Andriy H. Nevidomskyy$^{2}$}
\author{Elbio Dagotto$^{1,4}$}
\author{Qimiao Si$^{2}$}
\author{Hsin-Hua Lai$^{2}$}
\affiliation{
$^1$ Department of Physics and Astronomy, University of Tennessee, Knoxville, Tennessee 37996, USA\\
$^2$ Department of Physics and Astronomy \& Rice Center for Quantum Materials, Rice University, Houston, Texas 77005, USA\\
$^3$ Department of Physics, Cornell University, Ithaca, New York 14853, USA\\
$^4$ Materials Science and Technology Division, Oak Ridge National Laboratory, Oak Ridge, Tennessee 37831, USA
}

\maketitle

\section{Heuristic Derivation of the Generalized Widom Formula}
In $d$ dimensions we consider a specific real-space partition in which the boundary between the two subsystems is a plane whose normal direction is $\hat{n}_d$. This partition preserves the translational symmetries in $d-1$ dimensions perpendicular to $\hat{n}_d$, and one can perform partial Fourier transformation for all the physical degrees of freedom along these $d-1$ axes, since the momenta $k_{1,2,\cdots, d-1}$ are good quantum numbers. We thus view the momentum space as consisting of arrays of parallel 1D chains with spacings $\delta k_{1,2,...,d-1} = 2\pi/L_\bot$, where $L_\bot$ is the linear size of these transverse directions.

Using well-established results for free fermions and coupled harmonic lattice systems with critical surfaces \cite{Wolf, GioevKlich, dingprx12, Lai_EE4EBL}, we assume that each 1D chain in momentum space intersecting the critical Fermi surfaces (critical points) contributes a 2nd Renyi entanglement entropy (REE) $(c_{eff}/4) \ln \mathcal{L}_\|$ [or a von Neumann EE ($\nu$EE) $(c_{eff}/3) \ln \mathcal{L}_\|$] to the total leading REE~\cite{Calabrese2004, Calabrese2009}, where $\mathcal{L}_\|$ is the linear size of the (smaller) subsystem along $\hat{n}_d$ and $c_{eff}$ represents the effective number of free gapless modes for each 1D chain. Note that we only consider the universal part of the leading terms in REE \cite{Calabrese2004, Calabrese2009}, \textit{i.e.}, for each 1D chain the leading REE explicitly should be $\frac{c_{eff}}{4} \ln \mathcal{L}_\parallel + c_2$ \cite{Calabrese2009}, where $c_2$ is a non-universal constant that we ignore. For free fermions or Fermi liquids (FL) \cite{GioevKlich, Zhang11, dingprx12, Lai_EEreconstruct}, $c_{eff} = c_{F}=1$; for coupled harmonic lattice models realizing the lattice version of the Exciton Bose liquid phase (EBL)~\cite{Paramekanti02}, $c^{EBL}_{eff} =2$~\cite{Lai_EE4EBL}. For the Gutzwiller-projected wave function in 2D, the $c_{eff}$ is not known. Nevertheless, the leading REE can be obtained by counting the total number of chains (in momentum space) intersecting the critical surface, which corresponds to the critical surface cross-sectional area divided by the $(d-1)$ dimensional area spacing between the chains, {\it i.e.,} $(2\pi/L_\bot)^{d-1}$. Explicitly, the leading universal part of REE is
\begin{eqnarray}
S_{dD} &\dot{=}& \frac{c_{eff}}{4} \ln \mathcal{L}_\|  \frac{1}{2}  \frac{\int_{\partial \Gamma} \left| d\hat{S}_\Gamma \cdot \hat{n}_d\right|}{(2\pi/L_\bot)^{d-1}} \label{Eq:Lai}\\
\nonumber & = & \frac{c_{eff}}{8}\ln \mathcal{L}_\| \left( \frac{L_\bot}{2\pi}\right)^{d-1} \frac{\int_{\partial \Gamma}\int_{\partial A} \left | d\hat{S}_\Gamma \cdot d\vec{S}_A\right|}{2L_\bot^{d-1}} \\
& = & \frac{c_{eff}}{16} \frac{\ln \mathcal{L}_\|}{(2\pi)^{d-1}} \int_{\partial A} \int_{\partial \Gamma} \left| d\vec{S}_A \cdot d\hat{S}_\Gamma \right|,\label{supp:eq_WGK}
\end{eqnarray}
where $\dot{=}$ represents the leading contribution. $\mathcal{L}_\|$ is the linear size of the (smaller) subsystem along $\hat{n}_d$. The factor $1/2$ in the first line is due to the over counting of the cross-section. In the second line, we rewrite $\hat{n}_d$ as a real-space partition surface integral (with $d\vec{S}_A$ being the corresponding oriented area element whose direction is along the local normal direction) divided by the partition surface area in $d-1$ dimensions, $2L_\bot^{d-1}$. $\int_{\partial \Gamma}$ represents the surface integral along the critical surface in momentum space (with $d\hat{S}_\Gamma$ being the corresponding oriented area element). While we arrived at Eq.~\eqref{supp:eq_WGK} by considering a special partition, it is actually the correct formula for the free-fermion state for arbitrary cuts \cite{GioevKlich} if we set $c_{eff}=1$ and replace $\mathcal{L}_\|$ by the generic linear size of the smaller subsystem. For the Gutzwiller-projected wave function, if we consider an $L_y$-legged chain system with infinite length along the $x$-axis, the total number of gapless modes is $2L_y -1$.  In real 2D systems, taking periodic boundary conditions along the $y$ direction, we expect $c_{eff}$ for each line in momentum space to be $c_{eff} = 2 - 1/L_y\big{|}_{L_y\rightarrow \infty} \rightarrow 2$.

\begin{figure*}
   \centering
   \includegraphics[width=\linewidth]{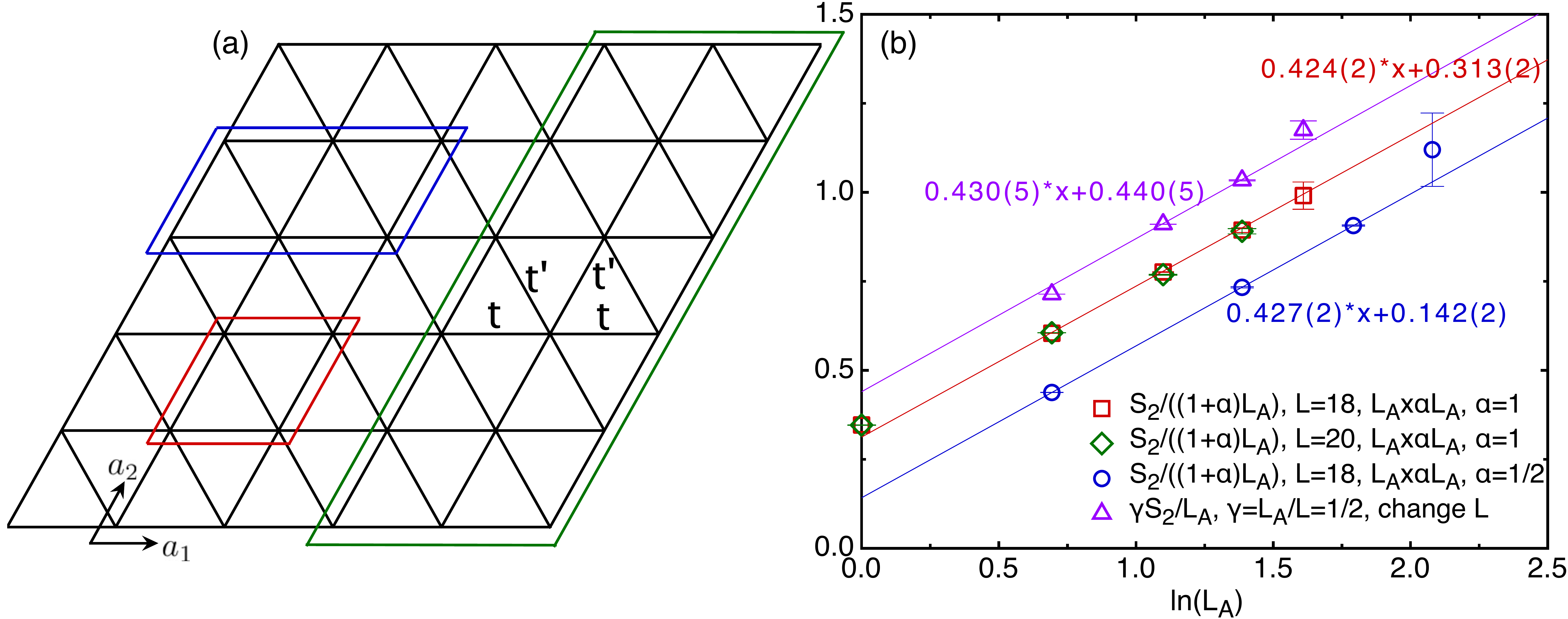}
   \caption{(a) Illustration of the different subsystem geometries on the triangular lattice that we consider for the REE calculations. We set $t\equiv 1$, and consider $t'/t = 1.0$ for the isotropic model. The colored regions represent the subsystems that we considered. The blue and red regions represent subsystems with size $L_A \times \alpha L_A$, where $\alpha =1/2, 1$. The green region represents the subsystem preserving translation along the $\bm{a}_2$ direction. (b) The REE, $S_2$, for different subsystem geometries for the isotropic case. We choose two setups to extract REE. For red squares, green diamonds and blue circles, we fix the whole system size and choose the subsystem size to be $L_A \times \alpha L_A$ with $\alpha = 1/2$ (blue circles), $1$ (red squares and green diamonds). For purple triangles, we fix the ratio of the linear size of the subsystem and that of the whole system to be $1/2$ ($L_A / L = 1/2$) and vary the value of $L$ to extract the REE. Based on Eq.~\eqref{supp:eq_WGK}, we plot $S_2/((1+\alpha)L_A)$ or $\gamma S_2/L_A$ vs $\ln (L_A)$. The slopes of the lines give the prefactor of the leading REE. }
\label{Fig:ee_fs_uni}
\end{figure*}

\section{REE by Variational Monte Carlo}
The variational wave function used here for a critical spin liquid with a spinon Fermi surface (SFS) is defined as
\begin{eqnarray}
|\psi\rangle=\mathcal{P}_{G}|\psi_{0}\rangle,
\end{eqnarray}
where $\mathcal{P}_{G}=\prod_{i}(1-n_{i\uparrow}n_{i\downarrow})$ is the Gutzwiller projector,
which enforces no double occupation on each site.
$|\psi_{0}\rangle$ is the ground state of the following mean-field Hamiltonian on the triangular lattice:
\begin{equation}\label{eq:SL}
{\cal H}_{\rm MF} = \sum_{\langle i,j\rangle,\sigma} t_{ij}c^{\dag}_{i,\sigma}c_{j,\sigma} + h.c.
\end{equation}
In a system with subsystems $A$ and $B$, the REE of order $n$ in $A$ is defined as
\begin{eqnarray}
S_{n}=\frac{1}{1-n}log[Tr\rho_{A}^{n}],
\end{eqnarray}
with the reduced density matrix associated with $A$, $\rho_{A}=Tr_{B}|\psi\rangle\langle\psi|$. We will focus on $n=2$ REE, $S_2$. To compute $S_2$, as in Ref.~\cite{Zhang11} we introduce an identical copy of the original system:
We divide the original system into two subsystems $a$ and $b$ and, likewise, the replica into $a'$ and $b'$. 
The operator $Swap$ is introduced as $Swap|a,b\rangle|a',b'\rangle=|a,b'\rangle|a',b\rangle$, and the REE is
\begin{eqnarray}
e^{-S_{2}}=\frac{\langle\Psi|Swap|\Psi\rangle}{\langle\Psi|\Psi\rangle},
\end{eqnarray}
with $|\Psi\rangle$ the wave function of the product between the original system and its replica. The operator
\begin{eqnarray}
\langle Swap\rangle=\underset{a,b;a',b'}{\sum}P(a,b,a',b')\frac{\psi(a',b)\psi(a,b')}{\psi(a,b)\psi(a',b')}
\end{eqnarray}
is calculated according to the weight $P(a,b,a',b')=\frac{|\psi(a,b)|^{2}|\psi(a',b')|^{2}}{\underset{a,b}{\sum}|\psi(a,b)|^{2}\underset{a',b'}{\sum}|\psi(a',b')|^{2}}$ by variational Monte Carlo (VMC). In this method, we generate the Markov chain for each configuration of both the original system and its replica, and use the Metropolis algorithm \cite{metropolis} to update the configurations according to the probability distribution $P(a,b,a',b')$. 

\begin{figure}
    \subfigure[von Neumann entanglement entropy]{\label{Fig:FFnuE_1to1}\includegraphics[width=\linewidth]{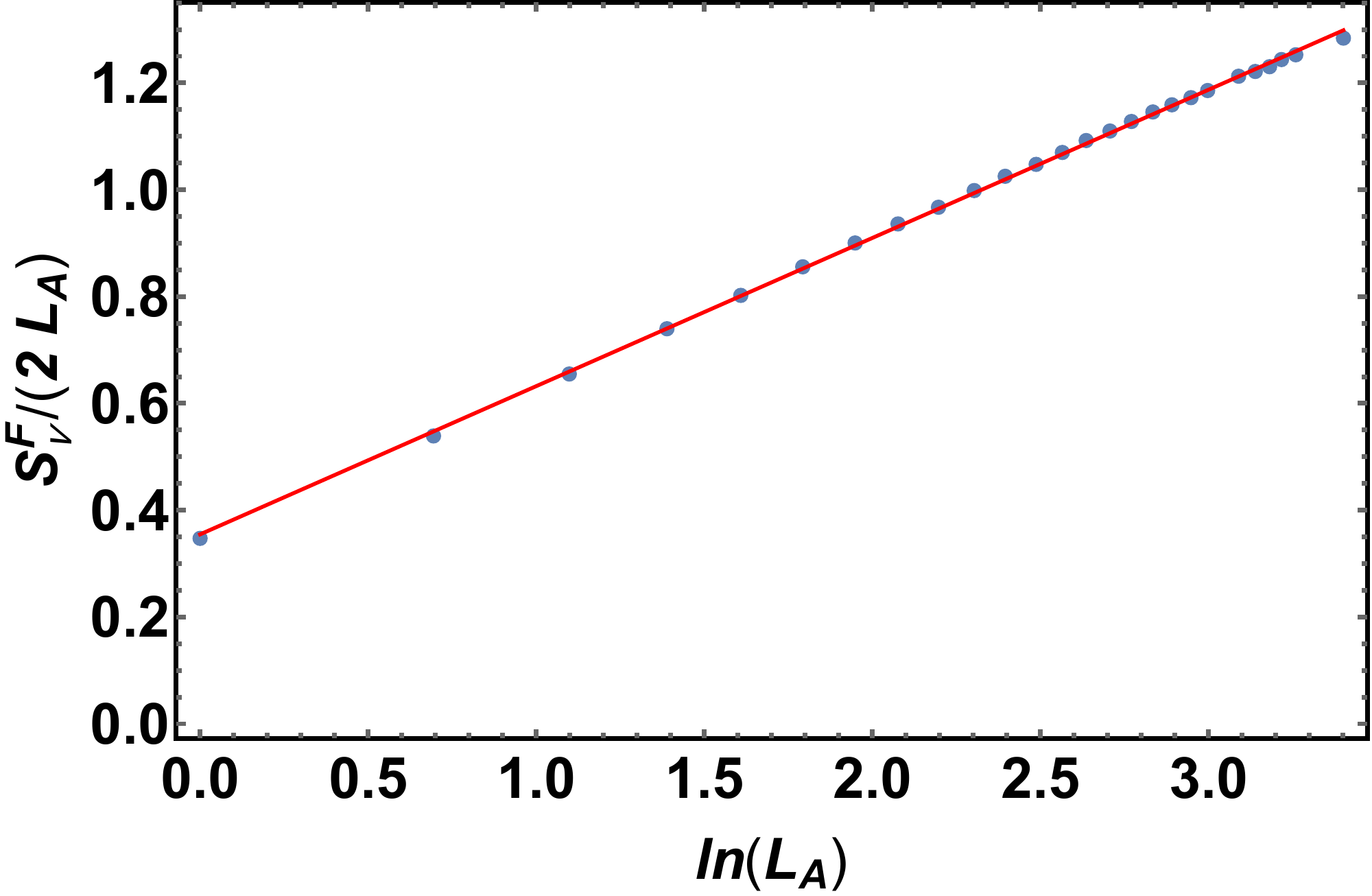}}
    \subfigure[Renyi entanglement entropy]{\label{Fig:FFRE_1to1}\includegraphics[width=\linewidth]{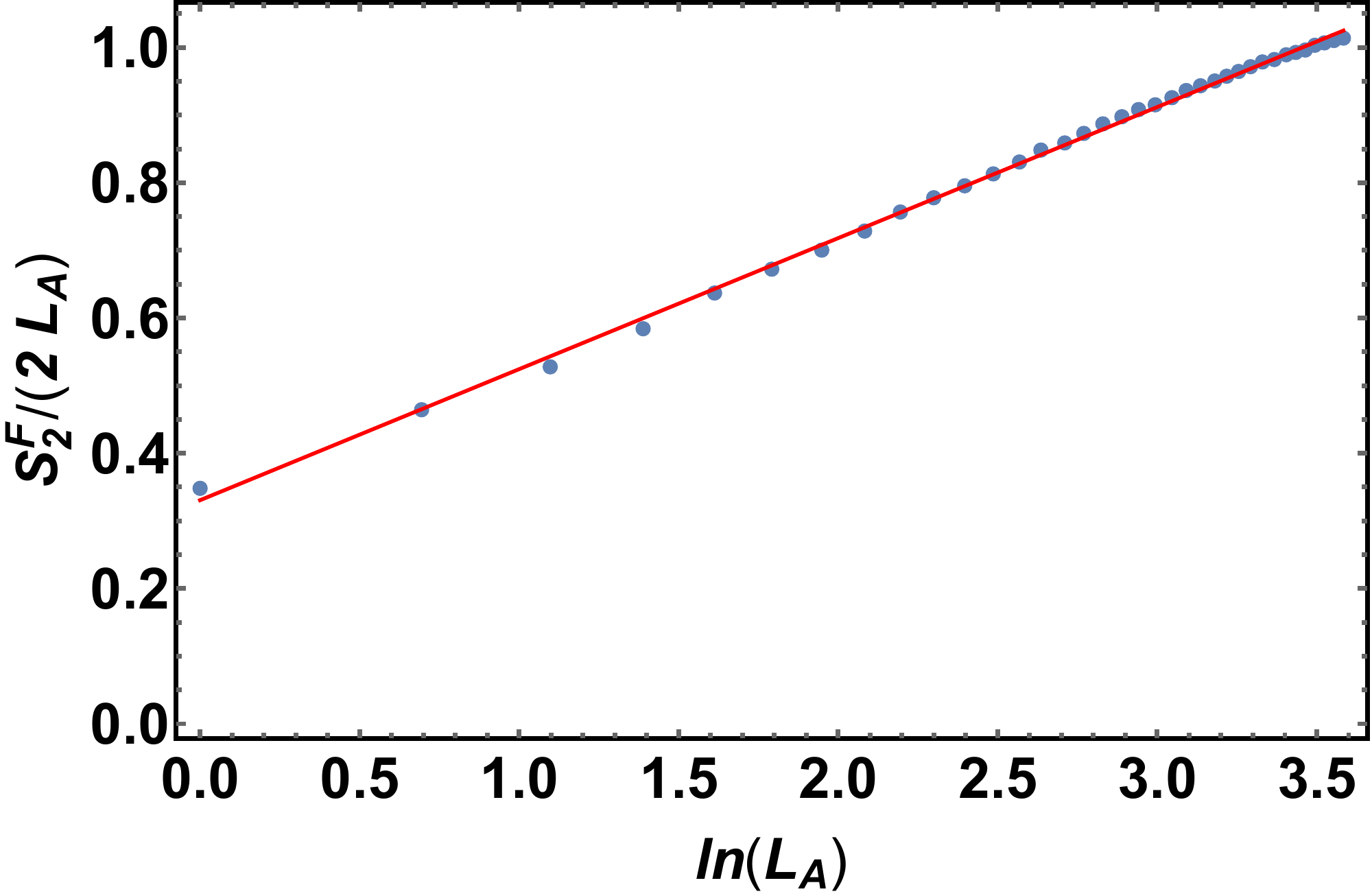}}
    \caption{The Fermi surface size of free fermions is fixed to be equal to the one obtained from the spin structure factor. Here we choose the subsystem with $L_A \times L_A$ lattice sites along the $\bm{a}_1$ and $\bm{a}_2$ directions and calculate the von Neumann entanglement entropy ($\nu$EE) (a) and the REE (b).  (a) We find that the fitting line is $y\simeq 0.278 x + 0.355$. (b) We observe a stronger oscillating behavior in the Renyi entropy, which makes it harder to obtain a conclusive fitting line. Based on the current data, we get the fitting line to be $y\simeq 0.194 x + 0.331$.}
\label{Fig:FFEE_121}
\end{figure}

\begin{figure}
    \subfigure[von Neumann entanglement entropy]{\label{Fig:FFnuE_bipartite}\includegraphics[width=\linewidth]{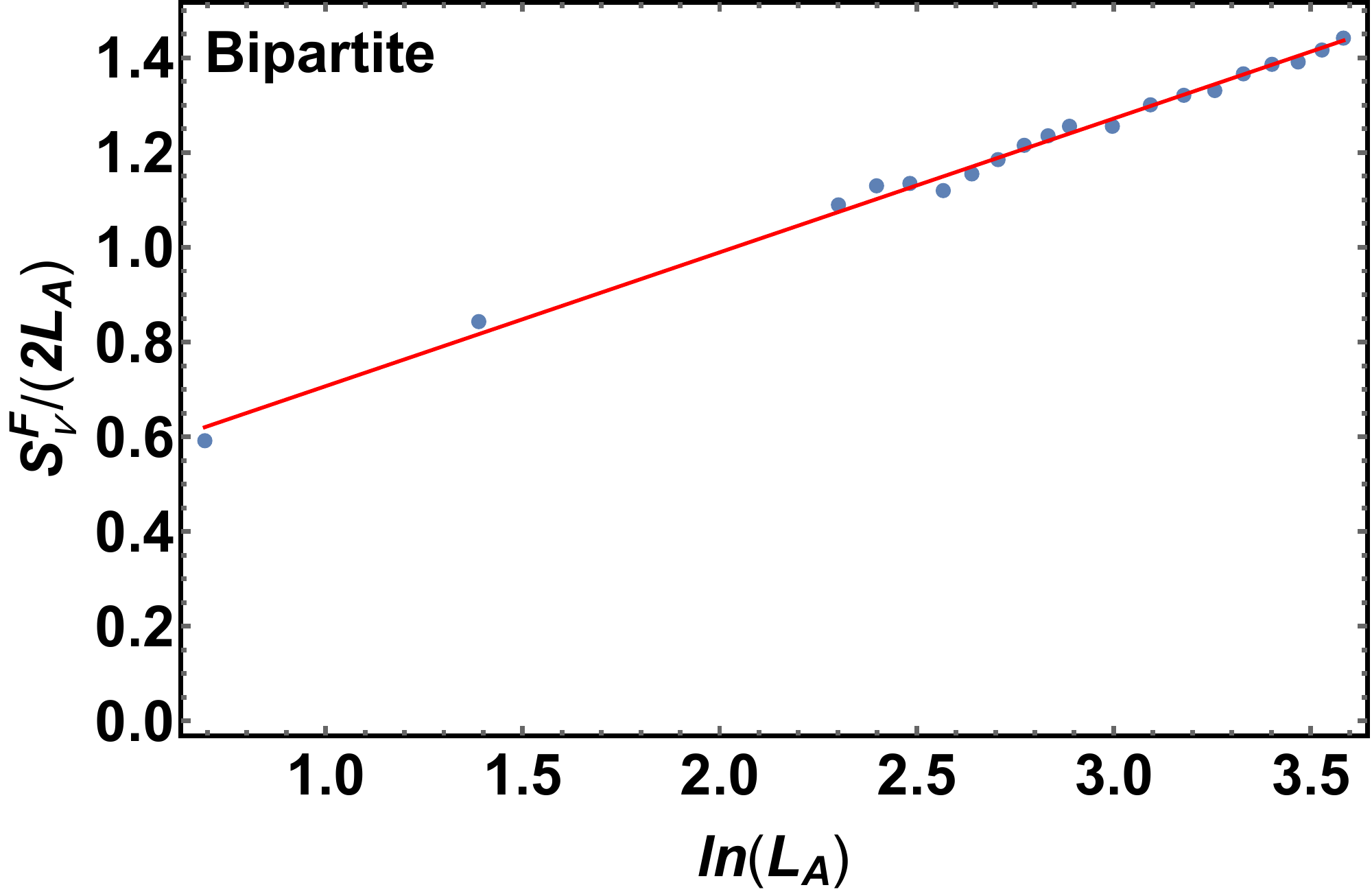}}
    \subfigure[Renyi entanglement entropy]{\label{Fig:FFRE_bipartite}\includegraphics[width=\linewidth]{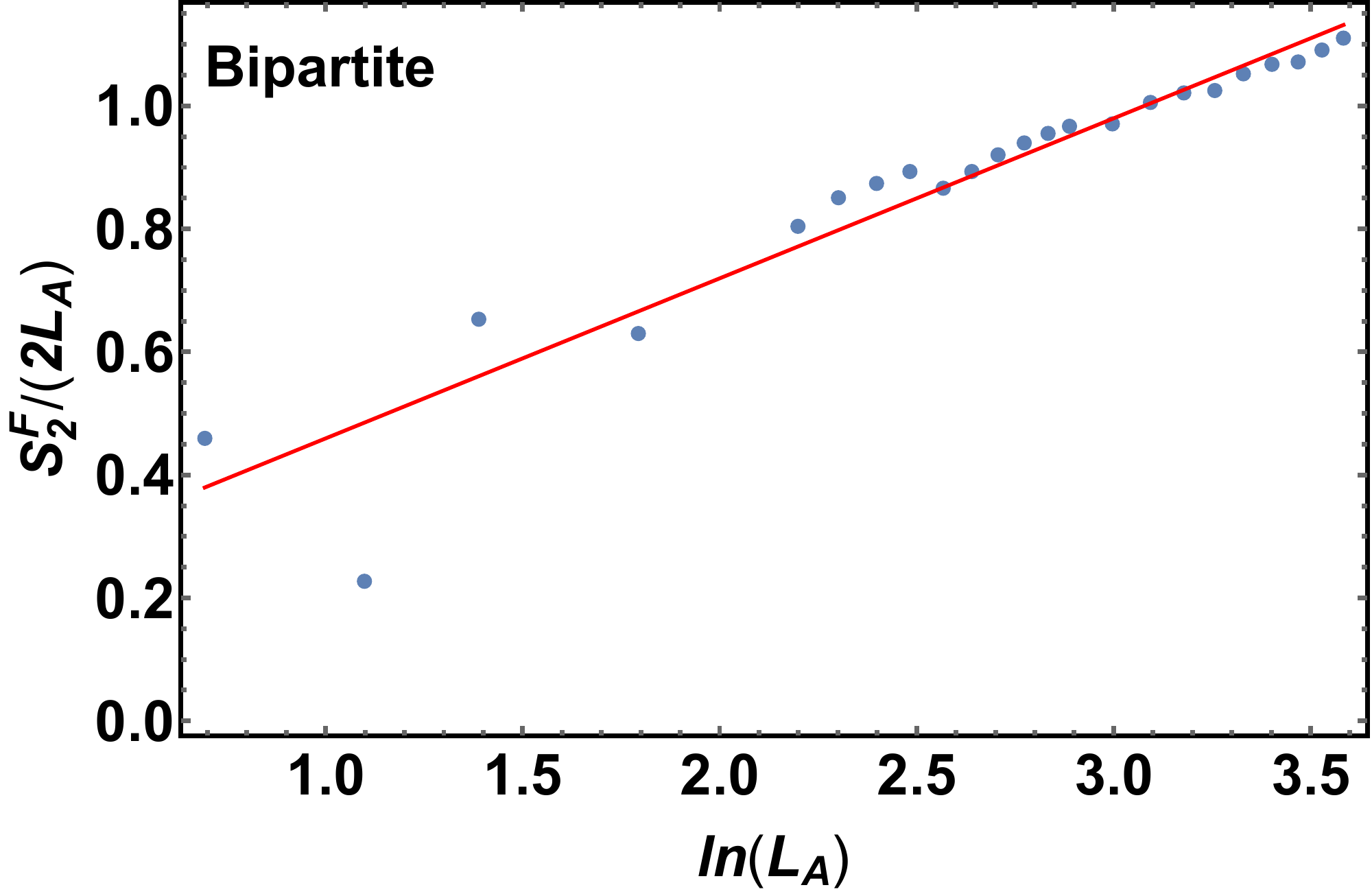}}
    \caption{The Fermi surface size of free fermions is fixed to be equal to the one determined from the spin structure factor result. Here we choose the subsystem to preserve the translation symmetry along the $\bm{a}_2$ direction to obtain the $\nu$EE (a) and REE (b). In both cases, we find stronger oscillating behaviors compared with those in Fig.~\ref{Fig:FFEE_121}, but we still can obtain a consistent result for the $\nu$EE. (a) We find the fitting line to be $y = 0.282 x + 0.424$, whose slope is comparable to that in Fig.~S\ref{Fig:FFnuE_1to1} obtained in a different subsystem geometry. (b) The REE shows a very strong oscillating behavior, which makes it very difficult to find a fitting line. A linear fitting line based on the current data is $y = 0.260 x + 0.199$.}
\label{Fig:FFEE_bipartite}
\end{figure}

\section{REE of the Isotropic Case by Variational Monte Carlo}
The results for REE in the isotropic case $(t=t')$, with $t/t'$ representing the fermion hopping amplitudes in the Hamiltonian~Eq.~(\ref{eq:SL}) (prior to Gutzwiller projection), are shown in Fig.~\ref{Fig:ee_fs_uni}. Panel (a) illustrates three different subsystems that we use for calculating the REE. The red and blue regions represent, in general, the subsystem with $L_A \times \alpha L_A$ sites, while the green region represents the subsystem with $L_A \times L$ sites with periodic boundary condition along the $\bm{a}_2$ direction, where $L$ represents the full system length. In panel (b), the slopes of the lines give the value of $c_{eff} A_{sf}$. The red squares are consistent with the data in Ref.~\cite{Zhang11} on a $18 \times 18$ triangular lattice with periodic boundary condition along both the $\bm{a}_1$ and $\bm{a}_2$ directions. The green diamonds represent our numerical data on a system with $20 \times 20$ sites along the $\bm{a}_1$ and $\bm{a}_2$ directions. We remark that the consistency between these two results shows the convergence of REE from $L=18$ to $L=20$ clusters, and then the $L=18$ cluster is sufficient to illustrate our VMC results. The open purple triangles represent the REE associated with the subsystem, which is half of the whole system in our setup, with periodic boundary conditions along $\bm{a}_2$. Within the error bars, we can see that the slope of the latter is \textit{equal} to the former obtained using a different subsystem geometry. The equality can be explained using the formula, Eq.~\eqref{supp:eq_WGK}, which we elaborate below.

In the first setup with $\alpha=1$, the REE is associated with the subsystem with four boundaries with equal surface area $L_A$ on a 2D triangular lattice with $L\times L$ sites. For each surface in real space, the integral along the real-space surface contributes $2 L_A A_{sf}$ to REE leading to $S^I_2\doteq c_{eff} A_{sf} /(4\pi)L_A \ln (L_A)$ based on Eq.~\eqref{supp:eq_WGK}. On the other hand, if the subsystem only preserves translational symmetry along the $\bm{a}_2$ direction, in real space there are only two surfaces with surface area $L$. For each boundary the integral gives $2 L A_{sf}$ that results in REE as $S^{II}_2 \doteq c_{eff} A_{sf} /(8\pi)L \ln (L_A) = (2\gamma)^{-1} c_{eff} A_{sf} /(4\pi) L_A \ln (L_A)$, where we rewrite $L = L_A/\gamma$. Setting $\gamma = 1/2$ gives $S^I_2 \doteq S^{II}_2$. The consistency between the leading terms of $S^{I}_2$ and $S^{II}_2$ as shown in Fig.~\ref{Fig:ee_fs_uni}(b) suggests the applicability of the Widom formula for the gapless spin liquid with a SFS described by the Gutzwiller-projected wave function. 

To further illustrate the applicability of the Widom formula to the gapless spin liquid states, we calculate the REE associated with a subsystem with different lengths along the $\bm{a}_1$ and $\bm{a}_2$ directions, which is chosen to be $L_A \times L_A/2$ in our work. If the Widom formula is applicable, the slopes of $S_2/((1+\alpha) L_A)$ vs $\ln(L_A)$ should be the same for different values of $\alpha$. The blue open circles in Fig.~\ref{Fig:ee_fs_uni}(b) are the data obtained in $L_A \times 1/2 L_A$ subsystems, and we can see that the slope is comparable to the previous results using different subsystem setups, which again suggest the validity of the Widom formula in this system.

\section{$\nu$EE and REE for Spinless Free Electrons in the Isotropic Case}
For free-fermion systems, we use the correlation function method \cite{Peschel2003} to obtain the $\nu$EE and REE in Figs.~\ref{Fig:FFEE_121} and \ref{Fig:FFEE_bipartite}. We use two kinds of subsystem setups illustrated in Fig.~\ref{Fig:ee_fs_uni}(a):  Setup 1: $L_A\times L_A$ subsystem [red region in Fig.~\ref{Fig:ee_fs_uni}(a)]; Setup 2: $L_A \times L$ subsystem with a periodic boundary along the $\bm{a}_2$ direction [green region in Fig.~\ref{Fig:ee_fs_uni}(a)]. The results are illustrated in Figs.~\ref{Fig:FFEE_121}-\ref{Fig:FFEE_bipartite}. In Fig.~\ref{Fig:FFEE_121}, we plot $S^F_\nu/((1+\alpha) L_A)$ $[S^F_2/((1+\alpha) L_A]$ vs $\ln(L_A)$, with $\alpha = 1$. In Fig.~\ref{Fig:FFEE_bipartite}, we plot $\gamma S^F_\nu/L_A$ $(\gamma S^F_2/L_A)$ vs $\ln(L_A)$, with $\gamma$ being the ratio between the subsystem length and that of the whole system, $\gamma = L_A/L = 1/2$. The numerical calculations have been performed up to the triangular lattice with $L=72$. In both cases, we observe stronger oscillating behaviors for the $S_2$ data in the free fermion systems, which makes it difficult to obtain the fitting lines. Focusing on the von Neumann entropy data, $S_\nu^F$, we find that the two different setups give comparable results. If we average the two slopes obtained in these two setups, we get the slope to be $\sim 0.28$. Utilizing the theoretical understandings that $S_\nu^F \dot{=} 4/3 S_2^F$ (the universal part of the entanglement entropy), we find that the theoretical slope for the REE data to be $0.21$, which is comparable to the average of the slopes in the two REE data in Figs.~\ref{Fig:FFEE_121}(b) and \ref{Fig:FFEE_bipartite}(b). The slope we obtained in this spinless free-fermion system is half of the value obtained in the Gutzwiller-projected wave function system, which again suggests that $c_{eff}\simeq 2$.

\begin{figure*}[t]
   \centering
   \includegraphics[width=\linewidth]{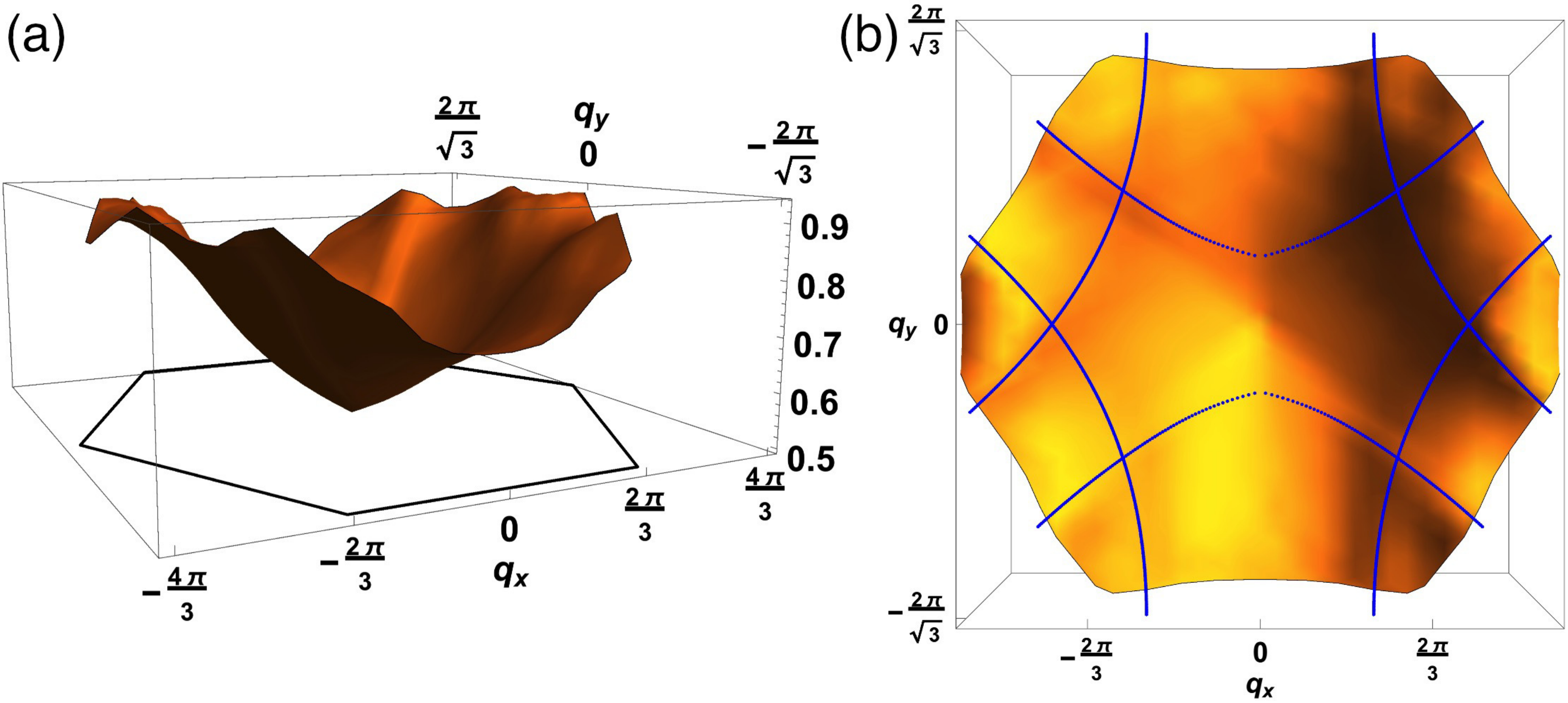}
   \caption{(a) Side view of the spin structure factor within BZ for the anisotropic case $t'/t = 0.7$, where the black hexagon represents the BZ. There is a sharp and singular peak at ${\bf q}=0$ corresponding to the uniform real-space power-law decaying behavior. The much weaker singular lines near the boundary of the BZ correspond to the oscillating real-space behavior caused by the presence of the SFS. (b) Top view of the spin structure factor. The weak singular lines on the surface are due to the presence of the emergent SFS and the theoretical locations of these lines are at $\bm{k}_{FR} - \bm{k}_{FL}$, \textit{i.e.}, for an observation direction we can define a momentum $\bm{k}_{FR/L}$ associated with a right-moving (left-moving) patch, represented by the blue lines. The theoretical blue lines match with remarkable accuracy  with the singular lines obtained by numerical calculations on a triangular lattice consisting of $30\times 30$ sites.
 }
   \label{Fig:sq_ani07}
 \end{figure*}

\section{Long Wavelength Analysis of the Spin Structure Factor}
To establish the precise formula for the leading entanglement entropy of the Gutzwiller-projected wave function describing the gapless spin liquid with a SFS, we can numerically determine $c_{eff}$ by comparing the result of the spin structure factor $\mathcal{D}_{\bf q}$ and the REE results, where
\begin{eqnarray}\label{Eq:Sq}
\mathcal{D}_{\bf q} \equiv \sum_j \chi^s_j e^{-i {\bf q} \cdot {\bf r}_j},
\end{eqnarray}
and $\chi^s_j $ is the real-space spin correlation functions defined as
\begin{eqnarray} \label{Eq:spincorr}
\chi^s_j \equiv \sum_{\mu =x, y, z} \la S^\mu_j S^\mu_{i=0}\ra.
\end{eqnarray}
The emergent SFS in momentum space can be viewed as consisting of patches of critical surfaces which at the low-energy descriptions are independent to each other \cite{SSLee2009}. It is expected that these independent patches contribute \textit{equally} to REE with the same $c_{eff}$. To determine $c_{eff}$, we focus on the isotropic critical surface case, \textit{i.e.}, a circular critical surface. For an arbitrary observing direction $\hat{n}$, we can determine two momenta corresponding to the right-moving patch, $\bm{k}^n_{FR}$, and the left-moving patch, $\bm{k}^n_{FL}$, respectively. For an isotropic convex critical surface with an inversion center, the vector ${\bf A}_s^n \equiv \bm{k}^n_{FR} - \bm{k}^n_{FL}$ must pass through the center of the critical surface and, therefore, the cross-section of the isotropic convex critical surface can be determined by the length $A_s^n = |{\bf A}_s^n|= |\bm{k}^n_{FR} -\bm{k}^n_{FL}|$. 

The wave vectors $\bm{k}^n_{FR} - \bm{k}^n_{FL}$ can be extracted by examining the spin structure factor. The power-law correlations in real space correspond to singularities in momentum space that can be revealed in the spin structure factor. At mean-field level, the power-law behavior can be explicitly determined. In general, the low-energy description dictates that the spin structure factor should show singularities at ${\bf q} = {\bf 0},~\bm{k}^n_{FR } - \bm{k}^n_{FL}$. Identifying $\bm{k}^n_{FR} - \bm{k}^n_{FL}$ along different directions can determine the cross-sections of the SFS at different directions. The exact result for the isotropic case is illustrated in Fig.~2 in the main text, which shows wiggling lines on the surface of the spin structure factor corresponding to the weak singularities of $\bm{k}^n_{FR} - \bm{k}^n_{FL}$. To numerically fit  the exact $\bm{k}^n_{FR} - \bm{k}^n_{FL}$, we assume that the Gutzwiller projection does not dramatically change the locations of the singularities, while only the exponents of the power-law behaviors of the singularities are modified. We then adopt a mean-field fermionic state with a Fermi surface at $1/2$-filling. Extracting $\bm{k}^n_{m,FR} - \bm{k}^n_{m,FL}$ of the mean-field ansatz, we find that the $\bm{k}^n_{m,FR} - \bm{k}^n_{m, FL}$ can fit the exact $\bm{k}^n_{FR} - \bm{k}^n_{FL}$, which indeed suggests that the Gutzwiller projection does not change dramatically the geometric information of the SFS in this system. 

At mean-field level, the Hamiltonian for an isotropic fermionic tight-binding model without enlargement of unit cells is
\begin{eqnarray}
H_{mf}  =  \sum_\alpha \sum_{\la {\bf r}, {\bf r}' \ra} f^{\alpha \dagger}({\bf r}) A({\bf r} - {\bf r}') f^\alpha ({\bf r}'),
\end{eqnarray}
where $\la...\ra$ represents nearest-neighbors and $\alpha = 1,2$ represents the spin flavor of fermions. $f^{\alpha \dagger}({\bf r}) (f^\alpha({\bf r}))$ represents the fermion creation (annihilation) operator with flavor $\alpha$ at location ${\bf r}$. We assume translational invariance so that $A_{{\bf r} {\bf r}'} = A({\bf r} - {\bf r}')$, where the matrix $A({\bf r} - {\bf r}')$ represents the hopping matrix between sites at ${\bf r}$ and ${\bf r}'$.  The mean-field Hamiltonian can be diagonalized by the complex fermions in Fourier space as
\begin{eqnarray}
f^\alpha({\bf r}) = \sqrt{\frac{1}{N_{s}}} \sum_{{\bf k}\in BZ} e^{i {\bf k}\cdot {\bf r}}\mathfrak{f}^{\alpha}({\bf k}),
\end{eqnarray}
where $N_{s}$ is the number of sites, and the complex fermion field $f$ satisfies the usual anti-commutation relations, $\{ \mathfrak{f}^{\alpha \dagger} ({\bf k}),\mathfrak{f}^{\alpha'}({\bf k}') \} = \delta^{\alpha \alpha'} \delta_{{\bf k} {\bf k}'}$. The component of the spin-$1/2$ spin operator is
\begin{eqnarray}
S^\mu({\bf r}) = \sum_{\alpha, \beta=1,2} f^{\alpha\dagger}({\bf r}) \left( \frac{\sigma^\mu_{\alpha \beta}}{2} \right) f^{\beta}({\bf r}),
\end{eqnarray}
which can be expressed in Fourier space as
\begin{eqnarray}
&& S^{\mu}({\bf r})\simeq \sum_{{\bf k},{\bf k'}\in {\bf B.Z.}}\sum_{\alpha,\beta}  \frac{\sigma^\mu_{\alpha \beta}}{2N_s} \mathfrak{f}^{\alpha \dagger}({\bf k}) \mathfrak{f}^\beta ({\bf k'}) e^{-i({\bf k}-{\bf k'})\cdot {\bf r}}.
\end{eqnarray}
In order to determine the long-distance behavior at separation ${\bm r}$, we focus on patches near the Fermi surface where the group velocity is parallel or antiparallel to the observation direction $\hat{\bm n} = {\bm r}/|{\bm r}|$, because at large separation $|{\bm r}| \gg k_F^{-1}$, the main contributions to the correlations arise precisely from such patches.  Specifically, we introduce Right(R) and Left(L) Fermi patch fields and the corresponding energies
\begin{eqnarray}\label{fermi patches:momentum space}
&& f^{\alpha,(\hat{\bm n})}_P(\delta {\bm k}) = \mathfrak{f}^{\alpha}({\bm k}^{(\hat{\bm n})}_{FP} + \delta {\bm k}) ~,\\
&& \epsilon^{(\hat{\bm n})}_P (\delta {\bm k}) = |{\bm v}^{(\hat{\bm n})}_{FP}| \left( P \delta k_\parallel + \frac{\mathfrak{C}^{(\hat{\bm n})}_P}{2} \delta k_\perp^2 \right) ~, \label{fermi patches:energy}
\end{eqnarray}
where the superscript $(\hat{\bm n})$ refers to the observation direction and $P = R/L = +/-$; ${\bm v}^{(\hat{\bm n})}_{FP}$ is the corresponding group velocity (parallel to $\hat{\bm n}$ for the Right patch and anti-parallel for the Left patch); $\mathfrak{C}_{P = R/L}$ is the curvature of the Fermi surface at the Right/Left patch; $\delta k_\parallel$ and $\delta k_\perp$ are respectively components of $\delta {\bm k}$ parallel and perpendicular to $\hat{\bm n}$.  It is convenient to define fields in real space
\begin{eqnarray}\label{fermi patches:real space}
f^{\alpha,(\hat{\bm n})}_P({\bm r}) \sim \sum_{\delta {\bm k} \in {\rm Fermi~Patch}} f^{\alpha,(\hat{\bm n})}_P(\delta {\bm k}) e^{i\delta{\bm k} \cdot {\bm r}} ~,
\end{eqnarray}
which vary slowly on the scale of the lattice spacing [from now on, we will drop the superscript $(\hat{\bm n})$].  In this long-wavelength analysis, the relevant terms in the spin operator are
\begin{eqnarray}\label{Eq:spincorr_longwl}
 S^{\mu}({\bf r})\sim \sum_{P,P'}\sum_{\alpha, \beta} \sigma^{\mu}_{\alpha \beta} f^{\alpha\dagger}_{P}({\bf r}) f^{\beta}_{ P'}({\bf r}) e^{-i({\bf k}_{FP}-{\bf k}_{FP'})\cdot {\bf r}}.
\end{eqnarray}
The above long-wavelength expression for the $S^{\mu}$ operator implies that the corresponding correlation function defined in Eq.~\eqref{Eq:spincorr} contains contributions with ${\bf q}={\bf 0}$ and  ${\bf q}_{-}\equiv\bm{k}_{FR}- \bm{k}_{FL}$. More explicitly, for a patch specified by $\epsilon_P (\delta {\bm k})$ in Eqs.~\eqref{fermi patches:momentum space}-\eqref{fermi patches:energy}, we can derive the Green's function for the continuum complex fermion fields as
\begin{eqnarray}\label{long-wavelength:green function}
\la f^{\alpha\dagger}_{R/L}({\bm 0}) f^{\alpha}_{R/L}({\bm r}) \ra = \frac{\exp[\mp i \frac{3\pi}{4}]}{2^{3/2} \pi^{3/2} \mathfrak{C}_{R/L}^{1/2} |{\bm r}|^{3/2}} ~.
\end{eqnarray}
Using this and Eq.~\eqref{Eq:spincorr_longwl}, we can obtain the spin correlation
\begin{eqnarray}
\chi_s({\bf r}) &\sim& -\frac{1}{\mathfrak{C}_{R} |{\bf r}|^3} - \frac{1}{\mathfrak{C}_{L}|{\bf r}|^3}~\label{uniform:spin}\\
&+& \frac{2 \sin [ ( \bm{k}_{FR}-\bm{k}_{FL})\cdot {\bf r}] }{\mathfrak{C}_{R}^{1/2}\mathfrak{C}_{L}^{1/2} |{\bf r}|^3}. ~\label{krminuskl:spin}
\end{eqnarray}
Focusing on the structure factors $\mathcal{D}_{\bf q}$ defined in Eq.~\eqref{Eq:Sq}, we expect that there should be a cone-shaped singularity at ${\bf q}=0$, based on Eq.~\eqref{uniform:spin}:
\begin{eqnarray}
\mathcal{D}_{{\bf q}\sim 0} \sim |{\bf q}|,
\end{eqnarray}
which can be seen straightforwardly by performing Fourier transform exactly or by scaling analysis with ${\bf q} \sim {\bf r}^{-1}$. Furthermore, the spin structure factor should also reveal the singular surface at ${\bf Q}_{-} $, as expected from Eq.~\eqref{krminuskl:spin}.  At the long wavelength analysis at the mean-field level, we note that the singularities are expected to be one-sided,
\begin{eqnarray}
&& \mathcal{D}_{{\bf Q}_{-} + \delta {\bf q}}\sim |\delta q_{||}|^{3/2} \Theta (-\delta q_{||}).\end{eqnarray}

Fitting the exact $\bm{q} = \bm{k}_{FR} - \bm{k}_{FL}$ in the spin structure factor data illustrated in Fig.~2(b) in the main text, we extract the cross sections of the emergent SFS in the isotropic case to be the $5.24 \pm 0.05$ (where we set the lattice constant $a$ to be 1 and $\hbar = 1$), which allows us to obtain $c_{eff}\simeq2.01 \pm 0.02$.

\section{Spin Structure Factor of the Anisotropic Case}
The Fig.~\ref{Fig:sq_ani07} shows the 3D spin structure factor obtained on a triangular lattice with $30 \times 30$ sites for the anisotropic case $t'=0.7t$. The spin structure factor is defined as $D_{\bf q} \equiv \sum_j \chi^s_j e^{-i {\bf q} \cdot {\bf r}_j}$ with the real-space spin correlation function $\chi^s_j \equiv \sum_{\mu =x, y, z} \la S^\mu_j S^\mu_{i \equiv0}\ra$. It is known that for an observation direction $\hat{n}$, $D_{\bm{q}}$ should show singular peaks at $\bm{q} = \bm{0},~\bm{k}^{\hat{n}}_{FR} - \bm{k}^{\hat{n}}_{FL}$ associated with forward and backward scattering process. The Fig.~\ref{Fig:sq_ani07}(a) gives a side view of the $D_{\bm{q}}$ in the hexagonal Brillouin zone (B.Z.) where we observe a sharp singular point at $\bm{q} = 0$ and the wiggle lines on the surface correspond to the singular lines located at $\bm{k}^n_{FR} - \bm{k}^n_{FL}$. Unlike the isotropic case, we can see that the structure factor breaks the $C_6$ rotation. The Fig.~\ref{Fig:sq_ani07}(b) shows the top view of $D_{\bm{q}}$. To determine the location of the $\bm{k}^n_{FR} - \bm{k}^n_{FL}$, we adopt a mean-field fermionic state with a Fermi surface at $1/2$-filling. Extracting $\bm{k}^n_{m,FR} - \bm{k}^n_{m,FL}$ of the mean-field ansatz, where the subscript $m$ means mean-field, we find that the $\bm{k}^n_{m,FR} - \bm{k}^n_{m, FL}$ can fit the exact $\bm{k}^n_{FR} - \bm{k}^n_{FL}$ quite well for an arbitrary observation direction $\hat{n}$, which gives the locations of the weak singular lines on the surface of the 3D spin structure factor. In Fig.~\ref{Fig:sq_ani07}(b) the blue lines are obtained by examining the mean-field SFS spin liquid state, which overlap the weak singular lines on the surface of the 3D spin structure factor obtained by exact numerical calculations.

\bibliography{supp_GW_EE}